\documentclass[twocolumn,showpacs,preprintnumbers,amsmath,amssymb]{revtex4}
\usepackage{bm}% bold math
\usepackage[colorlinks=true,linkcolor=blue,citecolor=blue,urlcolor=blue]{hyperref}% add hypertext capabilities
\usepackage{color}

\usepackage{graphicx}
\usepackage{epstopdf}

\begin{document}

%\preprint{APS/123-QED}

\title{Anomalous coercivity enhancement with temperature and tunable exchange bias in Gd and Ti co-doped BiFeO$_3$ multiferroics}
%\title{Study of N\'eel walls in soft magnetic hybrid structures}

%\title{Multiferroic properties and exchange bias effect in Gd and Ti co-doped BiFeO$_3$ at different temperatures }
%\title{Study of N\'eel walls in soft magnetic hybrid structures}

\author{Bashir Ahmmad}
\email[Author to whom correspondence should be addressed (e-mail): ]{arima@yz.yamagata-u.ac.jp}
\affiliation{Graduate School of Science and Engineering, Yamagata University, 4-3-16 Jonan, Yonezawa 992-8510, Japan.}

\author{M. Z. Islam}
\author{Areef Billah}
\author{M. A. Basith}
\email[Author to whom correspondence should be addressed (e-mail): ]{mabasith@phy.buet.ac.bd}
%\altaffiliation[Present address: ]{Department of Physics, Bangladesh University of Engineering and Technology, Dhaka-1000, Bangladesh.}
\affiliation{Department of Physics, Bangladesh University of Engineering and Technology, Dhaka-1000, Bangladesh.
}

\date{\today}% It is always \today, today,
             %  but any date may be explicitly specified

\begin{abstract}
	We have investigated the effect of temperature on magnetic properties of Bi$_{0.9}$Gd$_{0.1}$Fe$_{1-x}$Ti$_x$O$_3$ (x = 0.00-0.20) multiferroic system. Unexpectedly, the coercive fields ($H_{c}$) of this multiferroic system increased  with increasing temperature. The coercive fields and remanent magnetization were higher over a wide range of temperatures in  sample x = 0.10  i.e. in sample having composition Bi$_{0.9}$Gd$_{0.1}$Fe$_{0.9}$Ti$_{0.1}$O$_3$ than those of x = 0.00 and 0.20 compositions. Therefore, we have carried out temperature dependent magnetization experiments extensively for sample x = 0.10. The magnetic hysteresis loops at different temperatures exhibit an asymmetric shift towards the magnetic field axes which indicate the presence of exchange bias effect in this material system. The hysteresis loops were also carried out at temperatures 150 K and 250 K by cooling down the sample from 300 K in various cooling magnetic fields ($H_{cool}$). The exchange bias field ($H_{EB}$) values  increased with $H_{cool}$ and decreased with temperature. The $H_{EB}$ values were tunable by field cooling at temperatures up to 250 K. 
	
	%The observed dielectric anomaly of this sample around N\'eel temperature suggested a coupling between their magnetic and electric ordering.
	
%N\'eel walls in soft magnetic NiFe/NiFeGa hybrid stripe structures surrounded by a NiFe film are investigated by high resolution Lorentz transmission electron microscopic imaging. An anti-parallel orientation of magnetization in 1000 $nm$ wide neighboring unirradiated-irradiated stripes is observed by forming high angle domain walls during magnetization reversal. Upon downscaling the stripe structure size from $1000 \ nm$ to $200 \ nm$ a transition from a discrete domain pattern to an effective magnetic medium is observed for external magnetic field reversal. This transition is associated with vanishing ability of hosting high angle domain walls between adjacent stripes.   
   
\end{abstract}

%\keywords{Suggested keywords}%Use showkeys class option if keyword
                              %display desired
\maketitle
\section{Introduction} \label{I}
Multiferroic materials, in which ferromagnetic, ferroelectric, and/or ferroelastic orderings coexist, have attracted significant research interest since many years \cite{ref61,ref1,ref2,ref3, ref62} due to their potential applications in data storage media, spintronics and ferroelectric random-access memories. The co-existence of 'ferro'-orders in multiferroics allows the possibility that the magnetization can be tuned by an applied electric field and vice versa. Among the limited choices offered by the multiferroic materials, BiFeO$_3$ (BFO), one of the single-phase multiferroic materials at room temperature, exhibits the co-existence of ferroelectric ordering with Curie temperature (T$_C$) of 1123 K and antiferromagnetic (AFM) ordering with a N\'eel temperature (T$_N)$ of 643 K \cite{ref63}. In BiFeO$_3$, magnetic ordering is of antiferromagnetic type, having a spiral modulated spin structure (SMSS) with an incommensurate long-wavelength period of 62 nm \cite{ref63}. This spiral spin structure cancels the macroscopic magnetization and inhibits the observation of the linear magnetoelectric effect \cite{ref64}. These problems ultimately limit the use of bulk BiFeO$_3$ in functional applications.  Recent investigations demonstrated that in order to perturb the SMSS and improve the magnetic properties of BiFeO$_3$, partial substitution of Bi by rare-earth ions \cite{ref65,ref66}  or alkaline-earth ions \cite{ref67}  and also substitution of Fe by transition metal ions \cite{ref69, ref70} is an effective route. It is also reported that simultaneous minor substitution of Bi and Fe in BiFeO$_3$ by ions such as La and Mn, La and Ti, Nd and Sc etc., respectively \cite{ref71, ref72, ref74} also enhanced the magnetism and ferroelectricity in BiFeO$_3$. 

Recently, we have observed that simultaneous substitution of Gd and Ti in place of Bi and Fe, respectively in BiFeO$_3$ multiferroics improved their morphological, dielectric and magnetic properties at room temperature \cite{ref6}. Later on, another group also observed fascinating magnetic, optical and dielectric properties in this Gd and Ti co-doped BiFeO$_3$ ceramic system at room temperature \cite{ref73}. Therefore, in this investigation, we were interested to conduct experiments on temperature dependence of magnetic properties of Gd and Ti co-doped Bi$_{0.9}$Gd$_{0.1}$Fe$_{1-x}$Ti$_x$O$_3$ (x = 0.00-0.20)  multiferroic materials. In particular, our interest is  to investigate the effect of temperature on remanent magnetization, coercive fields and exchange bias (EB) fields of these bulk polycrystalline materials. Notably, the EB phenomenon which manifest itself by a shift of a magnetic hysteresis loop in systems containing ferromagnetic/antiferromagnetic bilayers has been studied extensively since many years due to its importance in spintronic applications \cite{ref7}. The EB effect usually occurs when the system is cooled down in an external magnetic field through the magnetic ordering temperatures. Most of the research in this field is focused on specially prepared systems, however, recent investigations reported the presence of EB effect in perovskite manganites \cite{ref8}, cobaltites \cite{ref9}, Heusler alloys \cite{ref75}  and multiferroics \cite{ref10}. Although EB has been observed in various bulk materials, the effect in most cases has been limited to far below room  temperature ($<$100 K)\cite{ref35, ref36}, thus making the systems less attractive for applications.

In our previous investigation, the magnetization vs. magnetic field ($M-H$) hysteresis loops indicated the presence of exchange bias effect in the Gd and Ti co-doped Bi$_{0.9}$Gd$_{0.1}$Fe$_{1-x}$Ti$_x$O$_3$ (x = 0.00-0.20) ceramics at room temperature although the biasing field was very small \cite{ref6}. Therefore, in the present investigation, we have explored exchange bias effect in this material system at different temperatures ranging from 20 K to 300 K. We have also investigated the influence of cooling magnetic fields at temperatures 150 K and 250 K to observe a tunable exchange bias in sample composition  Bi$_{0.9}$Gd$_{0.1}$Fe$_{0.9}$Ti$_{0.1}$O$_3$ (this composition is referred as sample x = 0.10 throughout the manuscript). 

%Finally, the temperature dependent dielectric properties of this material were investigated to observe the possibility of the presence of magnetic and electric ordering.   

%Now we are interested to investigate temperature dependent dielectric and magnetic properties of Gd and Ti co-doped Bi$_{0.9}$Gd$_{0.1}$Fe$_{1-x}$Ti$_x$O$_3$ (x = 0.00-0.20)  multiferroic ceramics. From temperature dependent magnetization experiments we intend to investigate exchange bias (EB) effect in this multiferroic system. 

\section{Experimental details} \label{II}
The polycrystalline samples having compositions Bi$_{0.9}$Gd$_{0.1}$Fe$_{1-x}$Ti$_x$O$_3$ (x = 0.00-0.20) were synthesized by using standard solid state reaction technique as was described in details in our previous investigation \cite{ref6}. The powder materials and ceramic pellets sintered at 1100 K were used for magnetic characterization. The $M-H$ hysteresis loops of Bi$_{0.9}$Gd$_{0.1}$Fe$_{1-x}$Ti$_x$O$_3$ (x = 0.00-0.20)  multiferroic ceramics were carried out at different temperatures using a Superconducting Quantum Interference Device (SQUID) Magnetometer (Quantum Design MPMS-XL7, USA). The temperature dependent magnetization measurements were carried out both at zero field cooling (ZFC) and field cooling (FC) processes \cite{ref11}. The magnetic hysteresis loops were also carried out at different temperatures by cooling down the sample from 300 K in various cooling fields ($H_{cool}$) \cite{ref11}.

 %The temperature dependent dielectric properties were measured using an impedance analyzer (Wayne Kerr) at different frequencies. 

\begin{figure}[hh]
	\centering
	\includegraphics[width=8cm]{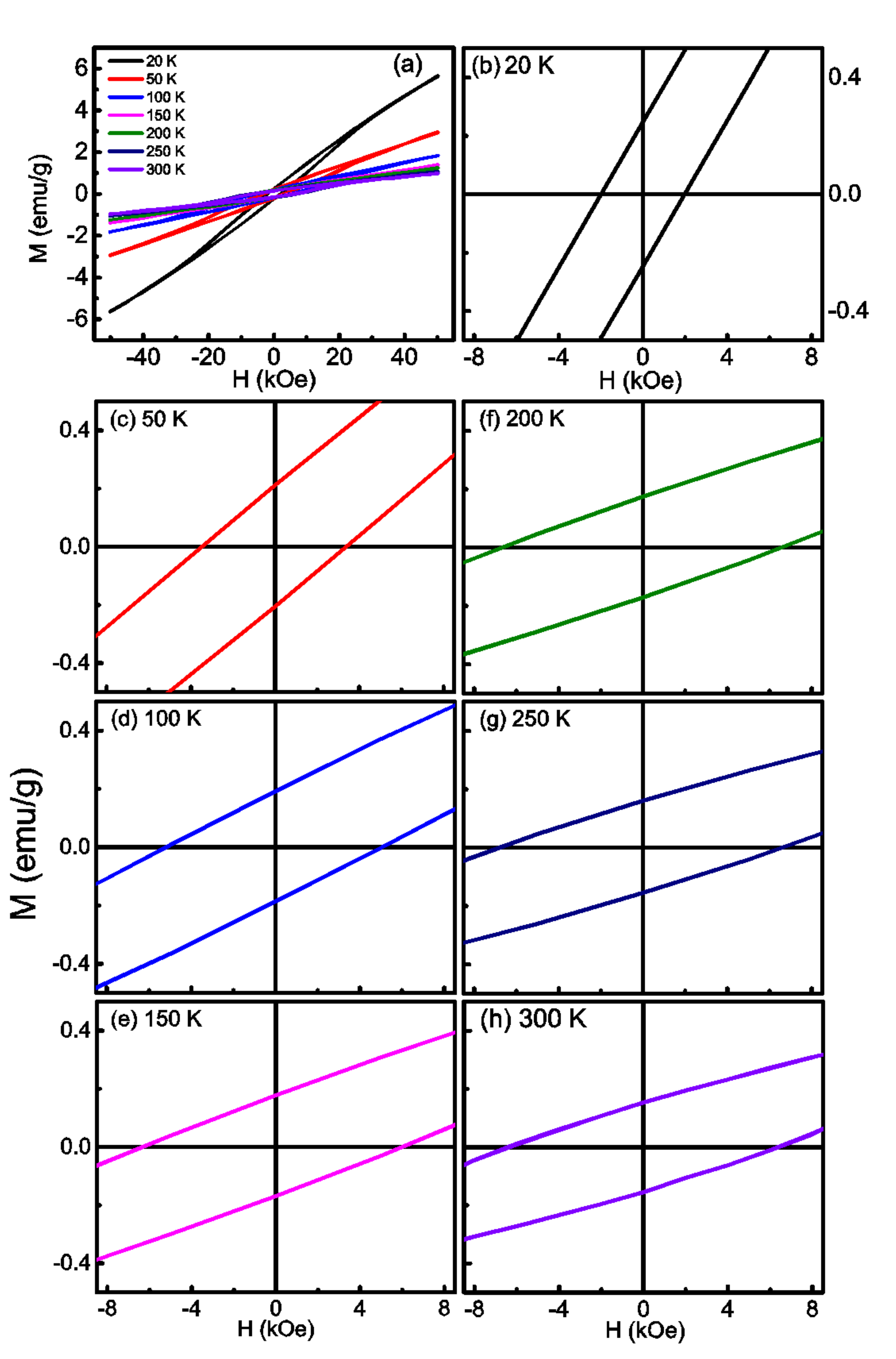}
	\caption{(a) The $M-H$ hysteresis loops of sample x = 0.10 (composition Bi$_{0.9}$Gd$_{0.1}$Fe$_{0.9}$Ti$_{0.1}$O$_3$) carried out at different temperatures. (b-h) An enlarged view of the low-field M-H hysteresis loops of sample x = 0.10 obtained at temperatures : (b) 20 K (c) 50 K (d) 100 K (e) 150 K (f) 200 K (g) 250 K and (h) 300 K.} \label{fig1}
\end{figure}

\section{Results and discussions} \label{III}
\subsection{Magnetic characterization} \label{I}
For magnetic characterization, the M-H hysteresis loops of Bi$_{0.9}$Gd$_{0.1}$Fe$_{1-x}$Ti$_x$O$_3$ (x = 0.00-0.20) samples were measured  at different temperatures ranging from 20 K to 300 K with an applied magnetic field of up to $\pm$50 kOe. As a typical example, the M-H loops of sample x = 0.10 (composition Bi$_{0.9}$Gd$_{0.1}$Fe$_{0.9}$Ti$_{0.1}$O$_3$) at different temperatures were presented in figure \ref{fig1} (a). Figures \ref{fig1} (b-h) demonstrate an enlarged view of the low-field M-H hysteresis loops of this sample measured at temperatures : (b) 20 K (c) 50 K (d) 100 K (e) 150 K (f) 200 K (g) 250 K and (h) 300 K. The hysteresis loops for the two other compositions were shown in the supplemental figures 1(S1) and 2(S2) \cite{ref99}. It should be noted that the M-H hysteresis loops (figure \ref{fig1}, supplemental figures 1(S1) and 2(S2)) were carried out initially without applying any cooling magnetic field.

%As shown in figure \ref{fig1} and supplemental figures 1(S1) and 2(S2), the Gd and Ti co-doped Bi$_{0.9}$Gd$_{0.1}$Fe$_{1-x}$Ti$_x$O$_3$ (x = 0.00-0.20) samples show unsaturated magnetization loops even with an applied magnetic field of up to $\pm$50 kOe which confirm the basic antiferromagnetic nature of the compounds. Notably, pure BiFeO$_3$ possesses a very narrow hysteresis loop with a very small but non-zero remanent magnetization (0.0009 emu/g) and a coercive field of $\sim {~}$110 Oe at room temperature \cite{ref6}. Compared to pure BiFeO$_3$, the center of M-H loops of Bi$_{0.9}$Gd$_{0.1}$Fe$_{1-x}$Ti$_x$O$_3$ (x = 0.00-0.20) compounds are wider (as shown typically in the enlarged view of sample x = 0.10, figures \ref{fig1} (b-h)) which indicate the co-existence of weak ferromagnetic nature along with the antiferromagnetic nature of the compounds \cite{ref11}. 

\begin{figure}[hh]
	\centering
	\includegraphics[width=9cm]{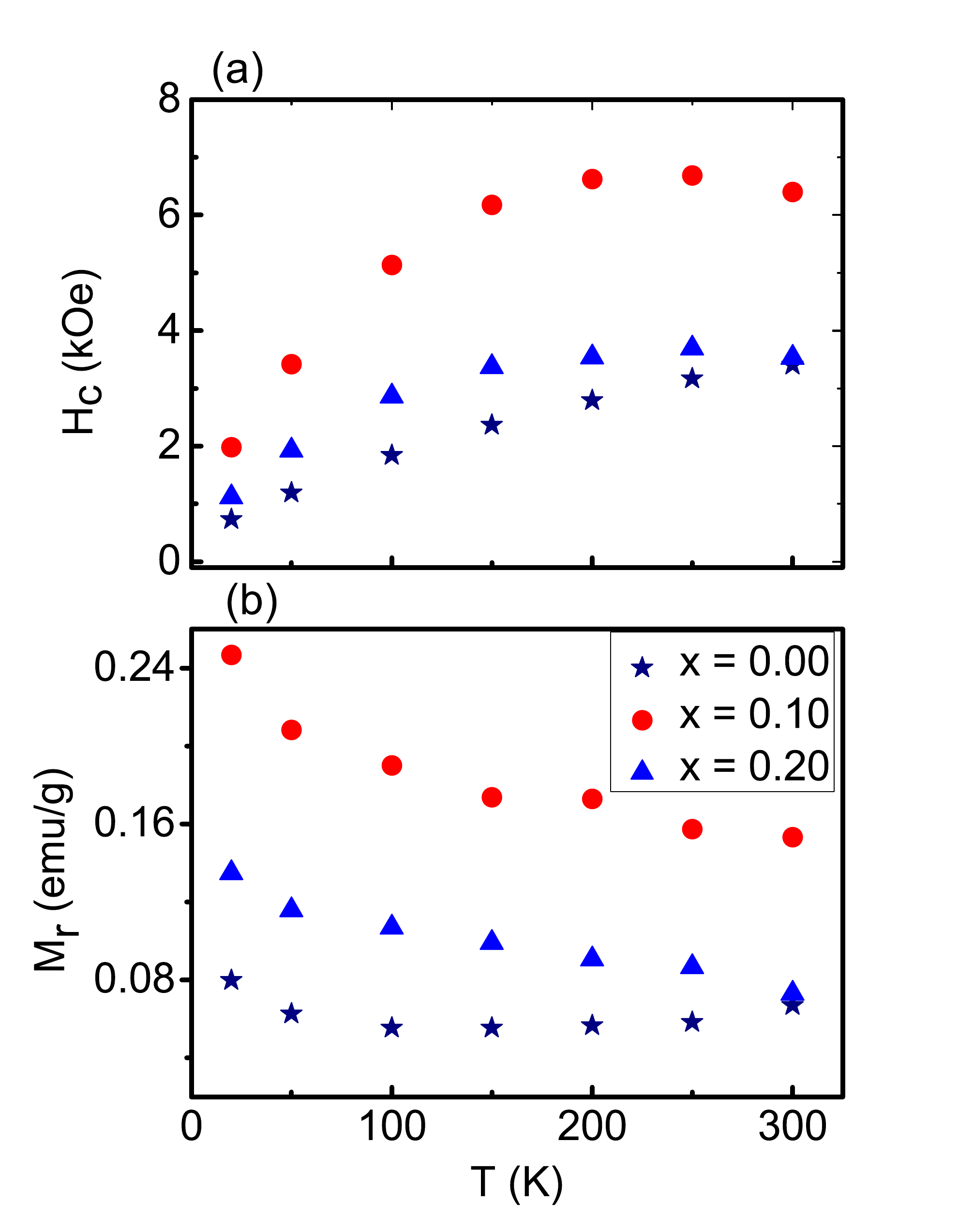}
	\caption{The variation of (a) coercivity (H$_c$) and (b) remanent magnetization (M$_r$) as a function of temperature in Bi$_{0.9}$Gd$_{0.1}$Fe$_{1-x}$Ti$_x$O$_3$ (x =0.00-0.20) compounds. Both H$_c$ and M$_r$ are higher over a wide range of temperatures in sample x =0.10.} \label{fig2}
\end{figure}

The coercive fields (H$_c$) and remanent magnetization (M$_r$) extracted from the hysteresis loops were quantified as: $H_c = (H_{c1}-H_{c2})/2$, where H$_{c1}$ and H$_{c2}$ are the left and right coercive fields \cite{ref6, ref12} and M$_{r}$ = $|$(M$_{r1}$-M$_{r2}$)$|$/2 where M$_{r1}$ and M$_{r2}$ are the magnetization with positive and negative points of intersection with H = 0, respectively \cite{ref71}. Calculated values of  H$_c$ and M$_{r}$ are plotted as a function of temperature in figures \ref{fig2} (a and b) respectively for Bi$_{0.9}$Gd$_{0.1}$Fe$_{1-x}$Ti$_x$O$_3$ (x =0.00-0.20) samples. Both the coercive fields and remanent magnetizations are higher for sample x = 0.10 than those for x = 0.00 and x = 0.20. Therefore, it is clear that the substitution of 10\% Ti in place of Fe in Bi$_{0.9}$Gd$_{0.1}$FeO$_3$ compound significantly increased H$_c$ and M$_{r}$. However, a further increment of Ti to 20\% in place of Fe reduced H$_c$ and M$_{r}$ although their net values are still higher than that of Ti undoped Bi$_{0.9}$Gd$_{0.1}$FeO$_3$ sample.  As was reported in our previous investigation \cite{ref6}, the larger values of H$_c$ and M$_r$ in sample x = 0.10 are related with the microstructure of the composition i.e. with homogeneous small grain size of the material. In the case of La and Nb co-substituted BiFeO$_3$ \cite{ref91} and Pr and Zr co-substituted BiFeO$_3$ compounds \cite{ref92}, increase in coercive field with the substitution was also attributed to decrease in grain size. The value of H$_c$ in these materials are much larger than that in pure BiFeO$_3$ \cite{ref6, ref74} and such a large value of coercivity in Gd and Ti co-substituted BiFeO$_3$ samples may be related to their magnetic anisotropy \cite{ref74,ref15,ref16}. 

\begin{figure}[!hh]
	\centering
	\includegraphics[width=9cm]{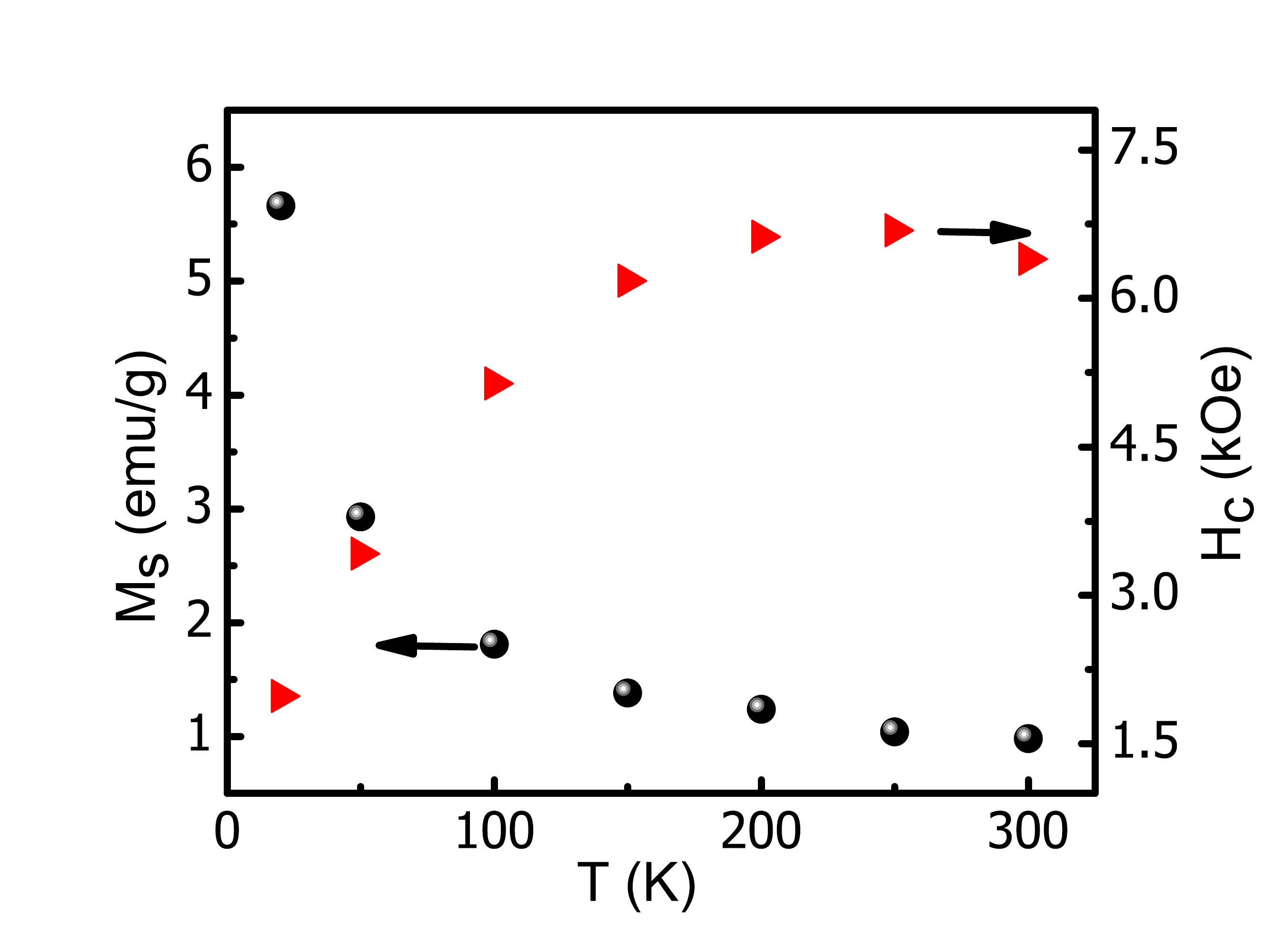}
	\caption{The variation of M$_s$ (the maximum magnetization at 50 kOe) and H$_c$ as a function of temperature in sample x = 0.10 (Bi$_{0.9}$Gd$_{0.1}$Fe$_{0.9}$Ti$_{0.1}$O$_3$).} \label{fig7}
\end{figure}

The remarkable feature observed from figure \ref{fig2}(a) is that the coercivity of these ceramic samples increases with temperature. Figure \ref{fig7} shows, as a typical example for x = 0.10 sample, that while the coercivity increases with increasing temperature, the maximum magnetization (M$_s$) at 50 kOe decreases with increasing temperature \cite{ref32}. For all studied samples, the coercivity show a strong temperature dependency and the H$_c$ values are significantly higher at room temperature than that at 20 K. For example, the H$_c$ value at room temperature for sample x = 0.10 is three times higher than that at 20 K. 

The usual trend in a typical magnetic system is for the coercivity to increase with decreasing temperature \cite{ref13, ref14,ref18} since the
anisotropy decreases much more sharply than does the magnetization with increasing temperature \cite{ref34}. In similar multiferroic material systems, e.g. La doped BiFeO$_3$  \cite{ref15,ref16} and LuFe$_2$O$_4$ \cite{ref17}, the H$_c$ values were found to increase at low temperature than that at room temperature. The unexpected decreasing trend of H$_c$ at low temperature as compared with the room temperature values for all compositions investigated here may be explained by the presence of magnetoelectric coupling in these multiferroic materials \cite{ref34, ref32}. The presence of the magnetoelectric coupling produces additional contribution to the anisotropy that actually acts to decrease the effective magnetic anisotropy \cite{ref34, ref32}. If $K_{u}$ is the uniaxial anisotropy constant in the
absence of coupling effects and  ${K_{u}^{'}}$ is the uniaxial anisotropy constant in the presence of magnetoelectric coupling, then   

$\mathrm{K_{u}^{'} = K_{u} - \chi_{\perp} \frac{(\beta P_z)^2}{2}}$
\\
\\
Here $\beta$ is the homogeneous magnetoelectric co-efficient that is related to the Dzyaloshinsky Moriya magnetic field, $\chi_{\perp}$ is the magnetic susceptibility in the direction perpendicular to the antiferromagnetic vector, $P_z$ is the spontaneous electric polarization \cite{ref34, ref33}. Thus the temperature variation in H$_c$ is determined by the competition between the magnetic anisotropy and the magnetoelectric coupling \cite{ref34, ref32, ref33}.

\begin{figure}[!hh]
	\centering
	\includegraphics[width=9cm]{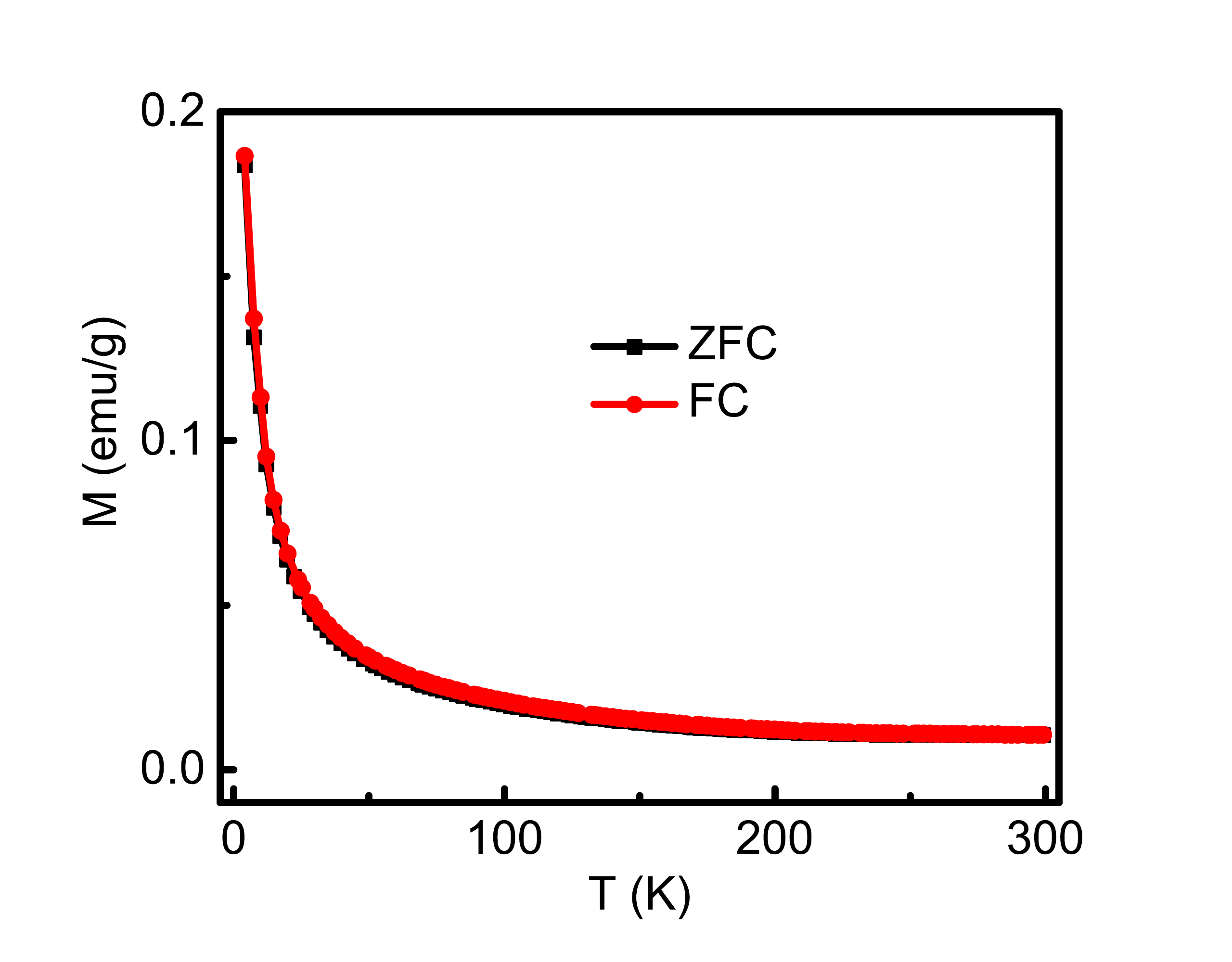}
	\caption{The temperature dependence of magnetization ($M-T$ curves) of Bi$_{0.9}$Gd$_{0.1}$Fe$_{0.9}$Ti$_{0.1}$O$_3$ sample measured in ZFC and FC processes in the presence of 500 Oe applied magnetic fields. Both ZFC and FC curves coincide with each other without showing any bifurcation.} \label{fig3}
\end{figure}

The higher values of H$_c$ and M$_r$ in sample x = 0.10 than those of x = 0.00 and 0.20 compositions motivated us to investigate further its magnetic properties in details. Therefore, we have carried out the temperature dependence of the magnetization ($M-T$) of Bi$_{0.9}$Gd$_{0.1}$Fe$_{0.9}$Ti$_{0.1}$O$_3$ bulk sample. The $M-T$ curves measured in ZFC and FC modes in the presence of 500 Oe applied magnetic field is shown in figure \ref{fig3}. In ZFC process, the sample is initially cooled from 300 K to the lowest achievable temperature and data were collected while heating in the presence of the applied field. On the other hand, in the FC mode, data values are collected while cooling in the presence of the magnetic field which is commonly known as ‘cooling magnetic field’ \cite{ref10}. Here the temperature dependence magnetization measurements demonstrate clearly that both ZFC and FC curves of Bi$_{0.9}$Gd$_{0.1}$Fe$_{0.9}$Ti$_{0.1}$O$_3$ ceramic coincide with each other without showing any bifurcation which indicates the absence of any spin flipping effect \cite{ref19, ref20, ref21}. In similar multiferroic materials both the ZFC and FC curves were also found to coincide with each other \cite{ref22, ref23}.
%In the ZFC process, the sample was initially cooled from 300 K to 5 K and data were collected while heating in the presence of the applied field. On the other hand, in the FC mode, data were collected while cooling in the presence of the magnetic field. 
\\

%The effect of the cooling magnetic field on exchange bias field and coercivity will be discussed later on.

%As shown in figure \ref{fig1} and supplemental figures 1(S1) and 2(S2), the Gd and Ti co-doped Bi$_{0.9}$Gd$_{0.1}$Fe$_{1-x}$Ti$_x$O$_3$ (x = 0.00-0.20) samples show unsaturated magnetization loops even with an applied magnetic field of up to $\pm$50 kOe which confirm the basic antiferromagnetic nature of the compounds. Notably, pure BiFeO$_3$ possesses a very narrow hysteresis loop with a very small but non-zero remanent magnetization (0.0009 emu/g) and a coercive field of $\sim {~}$110 Oe at room temperature \cite{ref6}. Compared to pure BiFeO$_3$, the center of M-H loops of Bi$_{0.9}$Gd$_{0.1}$Fe$_{1-x}$Ti$_x$O$_3$ (x = 0.00-0.20) compounds are wider (as shown typically in the enlarged view of sample x = 0.10, figures \ref{fig1} (b-h)) which indicate the co-existence of weak ferromagnetic nature along with the antiferromagnetic nature of the compounds \cite{ref11}.

As shown in figures 1(a-h), the room temperature $M-H$ hysteresis loop as well as the loops taken at other temperatures exhibit an asymmetric shift towards the magnetic field axes \cite{ref11}. This is a signature of the presence of an exchange bias effect in multiferroic Bi$_{0.9}$Gd$_{0.1}$Fe$_{0.9}$Ti$_{0.1}$O$_3$ material \cite{ref6, ref11, ref24}. In the present investigation, the hysteresis loops of Gd and Ti co-doped BiFeO$_3$ ceramic system is unsaturated even with an applied magnetic field of up to $\pm$50 kOe. which confirm the basic antiferromagnetic nature of the compounds. Notably, undoped BiFeO$_3$ possesses a very narrow hysteresis loop with a very small but non-zero remanent magnetization (0.0009 emu/g) and a coercive field of $\sim {~}$110 Oe at room temperature \cite{ref6}. Compared to pure BiFeO$_3$, the center of M-H loops of Bi$_{0.9}$Gd$_{0.1}$Fe$_{1-x}$Ti$_x$O$_3$ (x = 0.00-0.20) compounds are wider (as shown typically in the enlarged view of sample x = 0.10, figures \ref{fig1} (b-h)) which suggests a weak ferromagnetic nature \cite{ref81} of this co-doped ceramic system. The weak ferromagnetic nature of this material system is also revealed from temperature dependent magnetization curves \cite{ref81}. In this way we anticipate the co-existence of strong-anisotropic ferri/ferromagnetic (FM) and antiferromagnetic domains in this multiferroic material system. As a consequence of the exchange coupling at interfaces between these multiple magnetic domains, it is expected that the system acts as a natural system for generating EB effect in  Bi$_{0.9}$Gd$_{0.1}$Fe$_{0.9}$Ti$_{0.1}$O$_3$ multiferroics \cite{ref12, ref13, ref18, ref37}.
\begin{figure}[!t]
	\centering
	\includegraphics[width=9 cm]{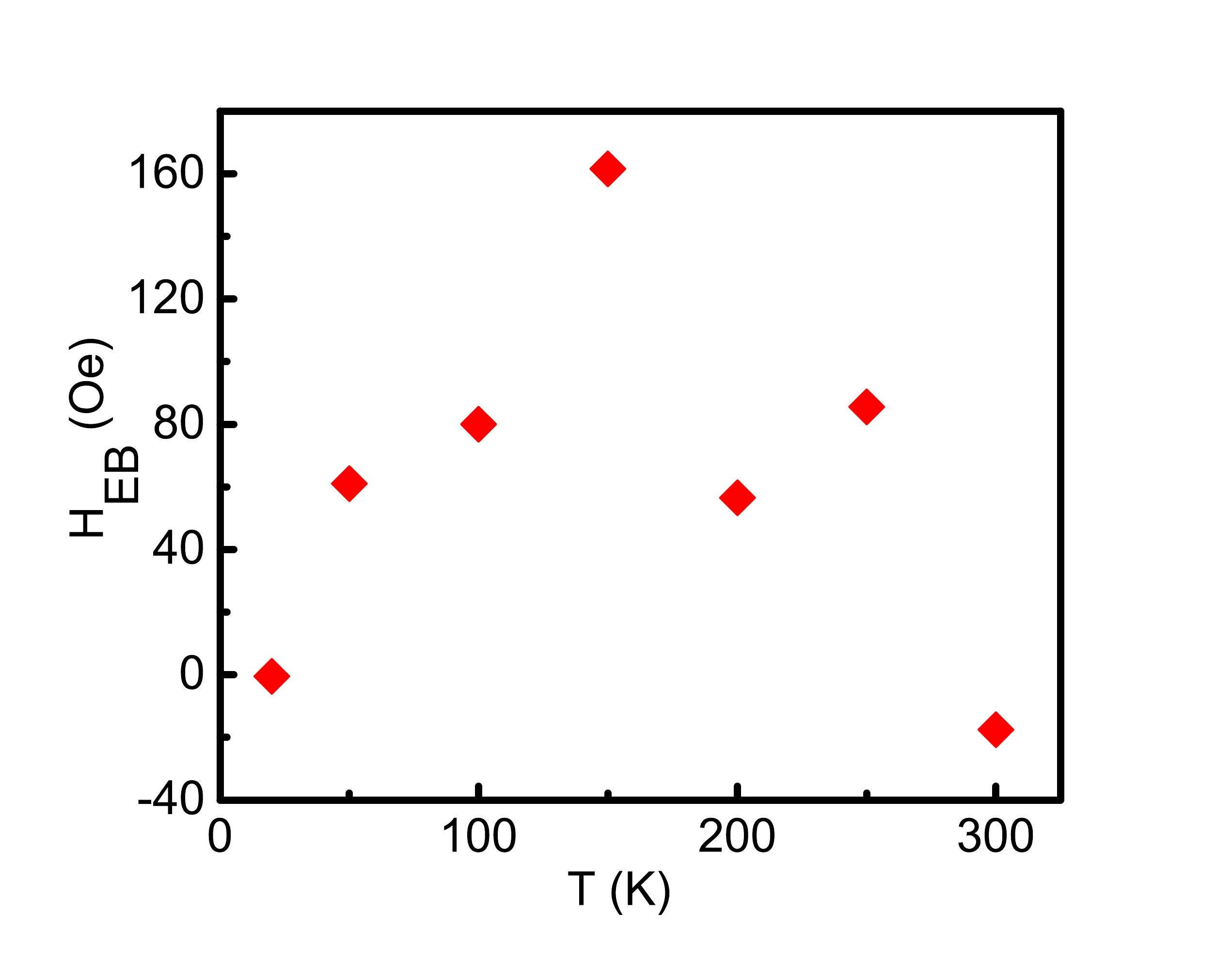}
	\caption{The variation of exchange bias fields (H$_{EB}$) as a function of temperature in Bi$_{0.9}$Gd$_{0.1}$Fe$_{0.9}$Ti$_{0.1}$O$_3$. The H$_{EB}$ values were calculated from the asymmetric shift of the $M$-$H$ hysteresis loops of figures \ref{fig1}(b-h). The hysteresis loops were taken without applying a cooling magnetic field.} \label{fig4}
\end{figure}
The exchange bias field (H$_{EB}$) from the loop asymmetry along the field axis can be quantified as $H_{EB} = -(H_{c1}+H_{c2})/2$ where $H_{c1}$ and $H_{c2}$ are the left and right coercive fields, respectively \cite{ref11,ref13}. The variation of H$_{EB}$ as a function of temperature in Bi$_{0.9}$Gd$_{0.1}$Fe$_{0.9}$Ti$_{0.1}$O$_3$ calculated from the asymmetric shift of the $M$-$H$ hysteresis loops of figures \ref{fig1}(b-h) is shown in figure \ref{fig4}. The temperature dependence of H$_{EB}$ for two other compositions were presented in the supplementary information (supplementary table 1) \cite{ref99}. 

As was mentioned earlier, the EB effect usually occurs when the system is cooled down in an external magnetic field through the N\'eel temperature (T$_N$). It is worth mentioning that in BiFeO$_3$ ceramic system the exchange bias effect was observed without any magnetic-field-annealing process through T$_N$ \cite{ref83} which is the conventional method of inducing unidirectional anisotropy \cite{ref84}.  It has also been observed without using any alloy layers \cite{ref85}, however, then the biasing strength of BiFeO$_3$ is observed to be very weak with  $H_{EB}$ = 36 Oe \cite{ref85} at room temperature. The exchange bias fields indicate the strength of the exchange coupling of an exchange bias system. As shown in figure \ref{fig4}, values of $H_{EB}$ of Bi$_{0.9}$Gd$_{0.1}$Fe$_{0.9}$Ti$_{0.1}$O$_3$ at 150 K and 250 K are higher than that at any other temperatures. The $H_{EB}$ values shown in figure \ref{fig4} is observed without applying any cooling magnetic field and therefore the biasing strength is weak and random. In the case of a FM/AFM bilayer system, it is well known that cooling magnetic field plays an important role in establishing a strong unidirectional anisotropy due to the exchange coupling \cite{ref80}. Hence, in the following experiments, the magnetic hysteresis loops of Bi$_{0.9}$Gd$_{0.1}$Fe$_{0.9}$Ti$_{0.1}$O$_3$ material were carried out at temperatures 150 K and 250 K by cooling down the sample from 300 K in various cooling magnetic fields ($H_{cool}$) ranging from 20 kOe to 60 kOe. In each experiment related to cooling magnetic fields, the measuring magnetic fields were from -30 kOe to 30 kOe. The details of the loop asymmetry at different cooling fields and temperatures can be found in the supplementary figures 3(S3) and 4(S4)  \cite{ref99}. 
	
%Therefore, we have investigated the influence of cooling magnetic fields on exchange bias fields of this multiferroic material system.	

\begin{figure}[!hh]
	\centering
	\includegraphics[width=9cm]{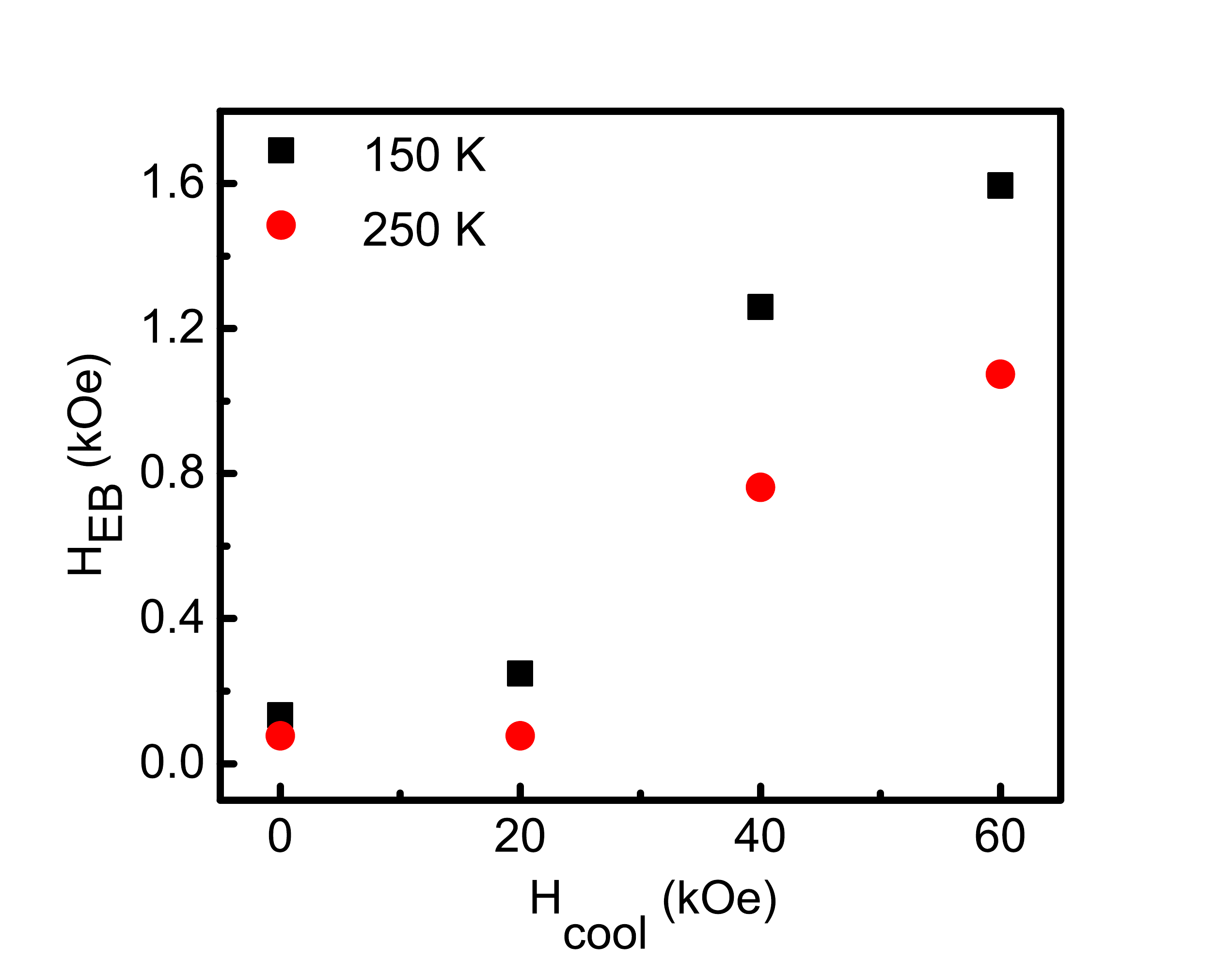}
	\caption{Cooling field dependence of exchange bias field ($H_{EB}$) at 150 k and 250 K for Bi$_{0.9}$Gd$_{0.1}$Fe$_{0.9}$Ti$_{0.1}$O$_3$ ceramic. The graph also demonstrates the dependency of $H_{EB}$ on temperature. } \label{fig5}
\end{figure}

%In the next stage of this investigation, we have investigated the exchange bias effect which were  manifested by an asymmetric shift of the $M$-$H$ curves of figures \ref{fig1}(b-h) measured at different temperatures.

%Notably, using the SQUID magnetometer (Quantum Design MPMS-XL7, USA) it was not possible to heat the sample above 300 K, therefore the field cooling experiments were not possible to conduct at 300 K.

The influence of cooling magnetic fields on exchange bias effect at temperatures 150 K and 250 K in Bi$_{0.9}$Gd$_{0.1}$Fe$_{0.9}$Ti$_{0.1}$O$_3$ multiferroic material is illustrated in figure \ref{fig5}. The $H_{EB}$ values increased significantly upon the application of cooling magnetic fields. For example, at temperature 250 K and cooling magnetic field 60 kOe, the biasing field increased more than twelve times than that for a cooling magnetic field of 20 kOe. During the field cooling experiments, the $H_{EB}$  values  reduced with increasing temperature, figure \ref{fig5}, which is similar to results reported in previous investigations \cite{ref10,ref26,ref27}. The $H_{EB}$ values are higher than those in related material systems \cite{ref10, ref25} specifically at high temperatures. In the present investigation, the temperature dependent ZFC and FC measurements (figure \ref{fig3}) demonstrate the absence of any bifurcation from 300 K down to 5 K and therefore, we had conducted the field cooling experiments by cooling the sample from 300 K down to 150 K and 250 K. By using the SQUID magnetometer (Quantum Design MPMS-XL7, USA), it was not possible to heat the sample above 300 K; therefore, the field cooling experiments were not possible to conduct by cooling down this sample through T$_N$ which is around 520$\pm 10$ K for Bi$_{0.9}$Gd$_{0.1}$Fe$_{0.9}$Ti$_{0.1}$O$_3$ ceramic measured from temperature dependent dielectric measurements (data not shown here). As was mentioned earlier, the EB has been observed in various bulk materials, however, the effect in most cases has been limited to low temperatures \cite{ref35, ref36}. The observation of EB up to room temperature is of interest from the perspective of practical applications. Therefore, in future we intend to conduct field cooling experiments at room temperature by cooling Bi$_{0.9}$Gd$_{0.1}$Fe$_{0.9}$Ti$_{0.1}$O$_3$ material through T$_N$ which is above the room temperature.

The influence of $H_{cool}$ on coercive fields were also measured at temperatures 150 K and 250 K and the results are displayed in table 1. Notably, in figures \ref{fig2} and \ref{fig7}, we have shown the effect of temperature on $H_{c}$ without applying any cooling magnetic field. Here in table 1, we have inserted the effect of cooling magnetic field on $H_{c}$ and the measurements were carried out at two different temperatures close to room temperature. The exchange bias fields are strongly affected by cooling magnetic field as shown in figure \ref{fig5}, however, the $H_{c}$ values (table 1) are influenced merely by a small extent. This is not unusual, in the case of NiFe/Co bilayers, similar effect was previously observed, i.e. the exchange bias field is strongly affected by cooling magnetic field whereas the $H_{c}$ is weakly influenced \cite{ref80}.
 %Although, the exchange bias fields are strongly affected by $H_{cool}$, figure \ref{fig5}, the coercive fields (table 1) are influenced merely by a small extent.
 \begin{table}[!h]
 	\caption{The table shows the effect of cooling magnetic fields on $H_{c}$ of Bi$_{0.9}$Gd$_{0.1}$Fe$_{0.9}$Ti$_{0.1}$O$_3$ sample at temperatures 150 K and 250 K. Notably, the effect of temperatures on $H_{c}$ of Bi$_{0.9}$Gd$_{0.1}$Fe$_{0.9}$Ti$_{0.1}$O$_3$ sample without applying cooling magnetic fields was inserted in figure \ref{fig2}. \label{Tab1}} 
 	\begin{center}
 		\begin{tabular}{|l|l|l|l|l|}
 			\hline
 			${T(k)}$&\multicolumn{4}{c|}{$H_c$ (kOe)}  \\
 			\cline{2-5}
 			&$H_{cool}$ = &$H_{cool}$ = &$H_{cool}$ =&$H_{cool}$ = \\
 			&0 KOe&20 kOe &40 kOe&60 kOe \\
 			\hline
 			150&$4.7$&$4.7$&$5.0$&$5.0$\\
 			\hline
 			250&$5.6$&$5.6$&$5.8$&$5.8$\\
 			\hline
 			\hline
 		\end{tabular}
 	\end{center}
 \end{table}
 
In our previous investigation \cite{ref11}, we have synthesized nanoparticles of the same composition i.e. of Bi$_{0.9}$Gd$_{0.1}$Fe$_{0.9}$Ti$_{0.1}$O$_3$ with a mean size of 40-100 nm directly from their bulk powder by using the sonication technique described in Ref. \cite{ref38}. In this specially prepared nanoparticle system, we have observed exchange bias effect and the magnitude of the exchange bias fields were also found to increase with cooling magnetic fields \cite{ref11}. Obviously, the cooling field dependence of $H_{EB}$ values of Bi$_{0.9}$Gd$_{0.1}$Fe$_{0.9}$Ti$_{0.1}$O$_3$ bulk system were weaker than those of specially prepared Bi$_{0.9}$Gd$_{0.1}$Fe$_{0.9}$Ti$_{0.1}$O$_3$ nanoparticles. For example, in both bulk and nanoparticle system the highest $H_{EB}$ values were observed at temperature 150 K by applying 60 kOe cooling magnetic field. At this temperature and cooling magnetic field, the $H_{EB}$ value of the bulk system is 20 \% less than that of nanoparticles having sizes 40-100 nm. This is worth noting as the preparation of bulk materials is comparatively easy and straightforward considering the multistep processing for the preparation of nanoparticles.      

\section{Conclusions} \label{II}

We have observed a strong influence of temperature on coercive fields and exchange bias fields of Gd and Ti co-doped BiFeO$_3$ multiferroics. The coercive fields of this multiferroic system were enhanced anomalously with increasing temperature. The anomalous enhancement of the coercivity with increasing temperature and particularly the observation of the high values of $H_{c}$ near/at room temperature may be of future use in potential applications where a coercivity stability at high temperature is crucially effective. We have also observed the presence of exchange bias effect due to the interface exchange coupling between FM and AFM domains of Gd and Ti co-doped BiFeO$_3$ material system. The magnitude of the exchange bias fields in bulk Bi$_{0.9}$Gd$_{0.1}$Fe$_{0.9}$Ti$_{0.1}$O$_3$ were also found to increase with cooling magnetic field. This magnetically tunable exchange bias obtained in bulk multiferroic Bi$_{0.9}$Gd$_{0.1}$Fe$_{0.9}$Ti$_{0.1}$O$_3$ material up to temperatures 250 K is promising, as most of the bulk materials show EB only far below room temperature. We anticipate that the presence of both exchange and magnetoelectric coupling in Bi$_{0.9}$Gd$_{0.1}$Fe$_{0.9}$Ti$_{0.1}$O$_3$ multiferroic material might be worthwhile for potential applications in novel multifunctional devices.

\section{Acknowledgements}
This work was supported by The world Academy of Sciences (TWAS), Ref.:14-066 RG/PHYS/AS-I; UNESCO FR: 324028567 and JSPS KAKENHI (Grant No. 26810117). A part of this work was conducted in The Institute for Molecular Science (IMS), supported by Nanotechnology Platform Program (Molecule and Material Synthesis) of the Ministry of Education, Culture, Sports, Science and Technology (MEXT), Japan. The authors thank to Mr. Motoyasu Fujiwara at the Institute of Molecular Science (IMS) for his assistance during SQUID measurement.

\end{document}